\documentclass[12pt]{amsart}
\usepackage{amscd}
\usepackage{mathrsfs}
\newtheorem{theorem}{Theorem}[section]

\theoremstyle{definition}

\theoremstyle{remark}
\newtheorem{remark}[theorem]{Remark}
\numberwithin{equation}{section}
\begin{document}

\def\CO{\mathbb C}
\def\RE{\mathbb R}
\def\vs{ \mathsf v}
\def\p{\par\noindent}
\def\G{\mathcal G}
\def\H{\mathscr H}

\title[The Schr\"odinger Equation with a Moving Point Interaction
]{The Schr\"odinger Equation with a Moving Point Interaction in Three
Dimensions}

\author{Andrea Posilicano}

\address{Dipartimento di Fisica e Matematica, Universit\`a
dell'Insubria, 
I-22100
Como, Italy}

\email{posilicano@uninsubria.it}
\thanks{2000 {\it Mathematics Subject Classification.} Primary 47B25,
47D08; Secondary 47D06, 81Q10}
\thanks{{\it Key words and phrases.} Point interactions, singular 
perturbations, unitary propagators.}

\begin{abstract} In the case of a single point interaction we improve, 
by using different methods, the existence theorem for the unitary
evolution generated by a
Schr\"odinger operator with moving point interactions obtained by
Dell'Antonio, Figari and Teta.
\end{abstract}

\maketitle

\begin{section}{Introduction} 
Let us denote by $L^2(\RE^3)$, with the usual scalar product 
$\langle\cdot,\cdot\rangle_2$ and corresponding norm $\|\cdot\|_2$, 
the Hilbert space of square 
integrable measurable functions on $\RE^3$. 
By $H^1(\RE^3)$ and by $H^2(\RE^3)$ we denote the usual
Sobolev-Hilbert spaces
$$
H^1(\RE^3):=\left\{\psi\in L^2(\RE^3)\,:\, \nabla\psi\in
L^2(\RE^3)\right\}\,,$$
$$H^2(\RE^3):=\left\{\psi\in
H^1(\RE^3)\,:\,\Delta\psi\in L^2(\RE^3)\right\}\,.
$$ 
Let $$H\equiv-\Delta:H^2(\RE^3)\to L^2(\RE^3)$$
be the self-adjoint operator giving the Hamiltonian of a
free quantum particle in $\RE^3$. For any $y\in\RE^3$ let us consider the
symmetric operator $H^\circ_y$ obtained by restricting $H$
to the subspace $$\{\psi\in H^2(\RE^3)\,:\, \psi(y)=0\}\,.$$ Such a
symmetric operator has defect indices $(1,1)$. Any of its self-adjoint
extensions different from $H$ itself describes a point
interaction centered at $y$. One has the following (see \cite{[AGHKH]},
section I.1.1 as regards $H_{\alpha,y}$ and see \cite{[T]} as regards 
$F_{\alpha,y}$)
\begin{theorem}\label{sing}
Any self-adjoint extension of $H_y^\circ$ different
from $H$ itself is given by
$$
H_{\alpha,y}:D(H_{\alpha,y})\to L^2(\RE^3)\,,
$$
\begin{align*}
&D(H_{\alpha,y}):=\\&\{\psi\in
L^2(\RE^3)\,:\,\psi(x)=\psi_\lambda(x)+\Gamma_{\alpha}(\lambda)^{-1}
\psi_{\lambda}(y)\,\G_\lambda(x-y)\,,\ \psi_\lambda\in
H^2(\RE^3)\}\,,
\end{align*}
$$
(H_{\alpha,y}+\lambda)\psi:=(H+\lambda)\psi_\lambda\,,
$$
the definition being $\lambda$-independent. Here $\alpha\in\RE$,
$$
\Gamma_{\alpha}(\lambda)=\alpha+\frac{\sqrt{\lambda}}{4\pi}
\,,\qquad
\G_\lambda(x)=\frac{e^{-\sqrt{\lambda}\,|x|}}{4\pi|x|}$$
and $\lambda>0$ is chosen in such a way that
$\Gamma_{\alpha}(\lambda)\not=0$. 
The kernel of the resolvent of $H_{\alpha,y}$ is given by 
$$
(H_{\alpha,y}+\lambda)^{-1}(x_1,x_2)=\G_\lambda(x_1-x_2)+\Gamma_{\alpha}
(\lambda)^{-1}\G_\lambda(x_1-y)\,\G_\lambda(x_2-y)\,.
$$
The quadratic form associated with $H_{\alpha,y}$ is
$$
F_{\alpha,y}:D(F_y)\times D(F_y)\to\RE\,,
$$
\begin{align*}
&D(F_y):=\\&\left\{\psi\in
H^1(\RE^3)\,:\,\psi(x)=\psi_\lambda(x)+q_\psi\G_\lambda(x-y)\,,\ 
\psi_\lambda\in H^1(\RE^3)\,,\ q_\psi\in\CO
\right\}\,,
\end{align*}
$$
(F_{\alpha,y}+\lambda)(\psi,\phi)=\langle\nabla\psi_\lambda,\nabla\phi_\lambda\rangle_2+\lambda\langle\psi_\lambda,\phi_\lambda\rangle_2+
\Gamma_{\alpha}(\lambda)\,\bar q_\psi q_\phi\,.
$$
Moreover the essential spectrum of $H_{\alpha,y}$ is purely absolutely 
continuous,
$$
\sigma_{ess}(H_{\alpha,y})=\sigma_{ac}(H_{\alpha,y})=[0,\infty)\,,
$$
$$\alpha<0\quad\Rightarrow\quad\sigma_{pp}(H_{\alpha,y})=
-(4\pi\alpha)^2\,,$$
$$
\alpha\ge 0\quad\Rightarrow\quad\sigma_{pp}(H_{\alpha,y})=\emptyset\,.
$$
\end{theorem}
\vskip 10pt\noindent
Suppose now that the point $y$ is not fixed but 
describes a curve $y:\RE\to\RE^3$, thus producing the family
of self-adjoint operators $H_{\alpha,y}(t)\equiv
H_{\alpha,y(t)}$. Then one is interested in showing 
that the time-dependent Hamiltonian
$H_{\alpha,y}(t)$ generates a strongly continuous unitary propagator
$U_{t,s}$. Note that both the operator and the form domain of 
$H_{\alpha,y}(t)$ are strongly time-dependent. This renders
inapplicable 
the known general theorems (see \cite{[Ka]}, \cite{[Pa]}) 
and such a generation
problem is not trivial. \par 
By exploiting the explicit form of
$H_{\alpha,y}(t)$ and in the case of several moving point interactions,
Dell'Antonio, Figari and Teta obtained in \cite{[DFT]} the following
(here we state their results in the simpler case of a single point
interaction)
\begin{theorem}\label{existence} Suppose that $$y\in C^3(\RE;\RE^3)\,,\qquad\psi\in
C_0^\infty(\RE^3)\,,\qquad\psi(y(s))=0\,.$$ 
Then there exist an unique
strongly continuous unitary propagator $$U_{t,s}:L^2(\RE^3)\to
L^2(\RE^3)$$ such that 
\begin{equation}
\left(i\frac{d}{dt}\,U_{t,s}\psi,\phi\right)_t=F_{\alpha,y}(t)
(U_{t,s}\psi,\phi)\label{cauchy}
\end{equation}
for all $\phi\in D(F_y(t))$.
\end{theorem}  
Here $(\cdot,\cdot)_t$ denotes the duality between
$D(F_{y}(t))$ and its strong dual. Moreover the solution
$\psi(t):=U_{t,s}\psi$ has a natural representation (see
\cite{[DFT]}, equations (14)-(21) for the details). \par 
In the introduction of \cite{[DFT]} the authors conjectured that
$U_{t,s}$ defines a flow on
$D(F_y(t))$ which is continuous with respect to the Banach topology
induced by the quadratic form $F_{\alpha,y}(t)$. \par Here we show, by
using different methods, that if $y\in C^2(\RE;\RE^3)$ then 
this is indeed the case and Theorem \ref{existence} above 
holds for any $\psi\in D(F_y(s))$ (see Theorem
\ref{theo} for the precise statements).  
\par
Our proof procedes in the following conceptually simple way. 
Noticing that the unitary map
$$T_t\psi(x):=\psi(x+y(t))$$ transforms the equation 
$$
i\,\frac{d\psi}{dt}(t)=H\psi(t)
$$
into the nonautonomous one
$$
i\,\frac{d\psi}{dt}(t)=H_\vs(t)\psi(t)\equiv(H+i\vs(t)\!\cdot\!\nabla)\,\psi(t)\,,\qquad \vs(t)\equiv\frac{dy}{dt}(t)\,,
$$
we consider the point perturbations (at $y=0$) of
$H_\vs$, where $\vs$ is a given, time-independent vector in $\RE^3$. 
We realize (see Theorem \ref{form}) that 
the form domains $D(F_{\vs})$ of such
singular perturbations $H_{\vs,\alpha}$ of $H_\vs$ are
$\vs$-independent. Indeed one has $D(F_{\vs})\equiv D(F_0)$, where $D(F_0)$
is the form domain of $H_{\alpha,y}$, $y=0$. This
allows, in the case the vector $\vs$ is time-dependent, the
application of Kisy\'nski theorem (see the Appendix) thus obtaining 
a strongly continuous unitary propagator 
$\tilde U_{t,s}$ which is also a strongly continuous propagator on
$D(F_0)$ with respect to the Banach topology induced by 
the quadratic form associated with
$H_{\alpha,y}$, $y=0$. Moreover  
$F_{\vs,\alpha}$, the quadratic form associated with
$H_{\vs,\alpha}$, and $F_{\alpha,0}$, the quadratic form associated with
$H_{\alpha,0}$, are related by the identity (see Theorem \ref{form} again) 
$$
F_{\vs,\alpha}=F_{\alpha,0}+Q_\vs\,,
$$
where $Q_\vs$ is the quadratic form associated
with the natural extension of $i\vs\!\cdot\!\nabla$ to $D(F_0)$ (see
Remark 2.4). This
allows us to show (see
Theorem \ref{theo}) that 
$$
U_{t,s}:=T^{-1}_t\tilde U_{t,s}T_s
$$
satisfies (\ref{cauchy}) for any $\psi\in D(F_{\alpha,y}(s))$ and is a
continuous flow from $D(F_{\alpha,y}(s))$ onto $D(F_{\alpha,y}(t))$. 
In the case $y\in C^3(\RE;\RE^3)$ we also
show that $U_{t,s}$ maps $\tilde D(H_{\alpha,y}(s))$ onto 
$\tilde D(H_{\alpha,y}(t))$, where 
$$\tilde D(H_{\alpha,y}(t)):=V_t D(H_{\alpha,y}(t))\,,\quad
V_t\psi(x):=e^{i\vs(t)\cdot x/2}\psi(x)\,.$$ 

\end{section}

\begin{section}{The operator $-\Delta+iL_\vs$ with a point interaction}
Let us consider the linear operator
$-\Delta+iL_\vs$, where
$$ L_\vs\psi:=\sum_{k=1}^{3}\vs_k\partial_k\psi\,,
\quad \vs\equiv(\vs_1,\vs_2,\vs_3)\in\RE^3\,.
$$
Since, for any $\epsilon>0$,
\begin{align*}
\|L_\vs\psi\|^2_2\le&
\sum_{1\le k,j\le 3}\left|\vs_k\vs_j\langle\partial^2_{kj}\psi,\psi\rangle_2\right|\le
3|\vs|^2|\langle\Delta\psi,\psi\rangle_2|\\
\le& \frac{3}{2}\,
|\vs|^2\left(\epsilon\|\Delta\psi\|^2_2+\frac{1}{\epsilon}\,\|\psi\|^2_2\right)\,,
\end{align*}
by Kato-Rellich theorem one has that
$$
H_\vs:=-\Delta+iL_\vs:H^2(\RE^3)\to L^2(\RE^3)
$$
is self-adjoint. Moreover, since, for any $\epsilon>0$,
\begin{align*}
|\langle L_\vs\psi,\psi\rangle_2|\le
|\vs|\,\|\nabla\psi\|_2\|\psi\|_2\le \frac{|\vs|}{2}\,
\left(\epsilon\|\nabla\psi\|^2_2+\frac{1}{\epsilon}\,\|\psi\|^2_2\right)
\,,
\end{align*}
$
H_\vs$ has lower bound $-|\vs|^2/4$.\par
Now we look for the self-adjoint extensions of the symmetric
operator $H_\vs^\circ$ obtained by restricting $H_\vs$ to the kernel of
the continuous and surjective linear map 
$$
\tau:H^2(\RE^3)\to\CO\,,\qquad \tau\psi:=\psi(0)\,.
$$
\begin{theorem} Any self-adjoint extension of $H_\vs^\circ$ different
from $H_\vs$ itself is given by
$$
H_{\vs,\alpha}:D(H_{\vs,\alpha})\to L^2(\RE^3)\,,
$$
$$
D(H_{\vs,\alpha}):=\left\{\psi\in
L^2(\RE^3)\,:\,\psi=\psi_\lambda+\Gamma_{\vs,\alpha}(\lambda)^{-1}
\psi_{\lambda}(0)\,\G^\vs_\lambda\,,\ \psi_\lambda\in H^2(\RE^3)\right\}\,,
$$
$$
(H_{\vs,\alpha}+\lambda)\psi:=(H_{\vs}+\lambda)\psi_\lambda\,,
$$
the definition being $\lambda$-independent. Here $\alpha\in\RE$,
$$
\Gamma_{\vs,\alpha}(\lambda)=\alpha+\frac{\sqrt{\lambda-|\vs|^2/4}}{4\pi}
\,,\qquad 
\G_\lambda^\vs(x)=
\frac{e^{-\sqrt{\lambda-|\vs|^2/4}\,|x|}}{4\pi|x|}\,e^{i\vs\cdot x/2}
$$
and $\lambda>|\vs|^2/4$ is chosen in such a way that
$\Gamma_{\vs,\alpha}(\lambda)\not=0$. 
The kernel of the resolvent of $H_{\vs,\alpha}$ is given by 
$$
(H_{\vs,\alpha}+\lambda)^{-1}(x_1,x_2)=\G^\vs_\lambda(x_1-x_2)
+\Gamma_{\vs,\alpha}(\lambda)^{-1}
\G^\vs_\lambda(x_1)\,\G^{-\vs}_\lambda(x_2)\,.
$$
Moreover  the spectrum of $H_{\vs,\alpha}$ is purely absolutely continuous, 
$$
\sigma_{ess}(H_{\vs,\alpha})=\sigma_{ac}(H_{\vs,\alpha})=[\,-|\vs|^2/4,\infty)\,,
$$
$$\alpha<0\quad\Rightarrow\quad\sigma_{pp}(H_{\vs,\alpha})=
-(4\pi\alpha)^2
-|\vs|^2/4$$
$$
\alpha\ge 0\quad\Rightarrow\quad\sigma_{pp}(H_{\vs,\alpha})=\emptyset\,.
$$
\end{theorem}
\begin{proof}
Let us define the bounded 
linear operators
$$
\breve G(\lambda):L^2(\RE^3)\to\CO\,,\qquad \breve G(\lambda)
:=\tau(H_\vs+\lambda)^{-1}$$
and
$$
G(\lambda):\CO\to L^2(\RE^3)\,,\qquad G(\lambda):=\breve
G(\lambda)^*\,.
$$
Since
$$
\G_\lambda^\vs(x)=\frac{e^{-\sqrt{\lambda-|\vs|^2/4}\,|x|}}{4\pi|x|}\,e^{i\vs\cdot x/2}$$ 
is the Green function of
$H_\vs+\lambda$ (see e.g. \cite{[O]}),  
one obtains
$$
\breve G(\lambda)\psi=\langle\G^{-\vs}_\lambda,\psi\rangle_2\,,\qquad
G(\lambda)q=q\,\G^\vs_\lambda\,.
$$ 
Since (see \cite{[P]}, Lemma 2.1)
$$
(\mu-\lambda)(H_{\vs}+\lambda)^{-1}G(\mu)
=G(\lambda)-G(\mu)\,,
$$
one obtains (see \cite{[P]}, Lemma 2.2)
\begin{align*}
&(\mu-\lambda)\breve G(\lambda)G(\mu)\\=&\tau(G(\lambda)-G(\mu))=
\tau(G(\nu)-G(\mu))-\tau(G(\nu)-G(\lambda))\\=&
\frac{\sqrt{\mu-|\vs|^2/4}}{4\pi}-\frac{\sqrt{\nu-|\vs|^2/4}}{4\pi}-
\left(\frac{\sqrt{\lambda-|\vs|^2/4}}{4\pi}-
\frac{\sqrt{\nu-|\vs|^2/4}}{4\pi}\,\right)\\=&
\frac{\sqrt{\mu-|\vs|^2/4}}{4\pi}-\frac{\sqrt{\lambda-|\vs|^2/4}}{4\pi}\,.
\end{align*}
The thesis about $H_{\vs,\alpha}$ and its resolvent then follows from Theorem
2.1 in \cite{[P]}. As regards the spectral properties of $H_{\vs,\alpha}$ one 
procedes as in \cite{[AGHKH]}, Theorem 1.1.4 .
\end{proof}
\begin{remark} Note that, as expected, $H_{\vs,\alpha}$ converges in norm
resolvent sense to $H_{\alpha,0}$ as $|\vs|\downarrow 0$.
\end{remark}
\begin{theorem}\label{form} The quadratic form associated with $H_{\vs,\alpha}$ is
$$
F_{\vs,\alpha}:D(F_0)\times D(F_0)\to\RE\,,\qquad
F_{\vs,\alpha}=F_{\alpha,0}+Q_\vs\,,
$$
where $D(F_0)$ is the domain of the quadratic form 
$F_{\alpha,0}$ associated with 
$H_{\alpha,y}$, $y=0$, (see
Theorem \ref{sing}) and
$$
Q_\vs:D(F_0)\times D(F_0)\to\RE
$$
$$
Q_\vs(\psi,\phi):=\langle iL_\vs\psi_\lambda,\phi_\lambda\rangle_2+
\bar q_\psi\langle \G_\lambda,iL_\vs\phi_\lambda\rangle_2
+q_\phi\langle iL_\vs\psi_\lambda,\G_\lambda\rangle_2\,. 
$$
\end{theorem}
\begin{proof}
Given $\psi$ and $\phi$ in $D(H_{\vs,\alpha})$ put 
$$
q_\psi:=\Gamma_{\vs,\alpha}(\lambda)^{-1}\psi_\lambda(0)\,,\qquad
q_\phi:=\Gamma_{\vs,\alpha}(\lambda)^{-1}\phi_\lambda(0)\,.
$$
Then
\begin{align*}
\langle (H_{\vs ,\alpha}+\lambda)\psi,\phi\rangle_2=&
\langle (H_{\vs}+\lambda)\psi_\lambda,\phi_\lambda\rangle_2
+q_\phi\langle (H_{\vs}+\lambda)\psi_\lambda,\G^\vs_\lambda\rangle_2\\
=&\langle (H_{\vs}+\lambda)\psi_\lambda,\phi_\lambda\rangle_2
+\Gamma_{\vs,\alpha}(\lambda)\,\bar q_\psi q_\phi\,.
\end{align*}
Thus one is lead to define the quadratic form
$$
F_{\vs,\alpha}:D(F_\vs)\times D(F_\vs)\to\RE
$$
by
$$
D(F_\vs):=\left\{\psi\in
H^1(\RE^3)\,:\,\psi=\psi_\lambda+q_\psi\G^\vs_\lambda\,,
\quad \psi_\lambda\in H^1(\RE^3)\,,\ q_\psi\in\CO
\right\}\,,
$$
\begin{align*}
&(F_{\vs,\alpha}+\lambda)(\psi,\phi)\\:=&
\langle(-\Delta+iL_\vs+\lambda)^{1/2}\psi_\lambda,
(-\Delta+iL_\vs+\lambda)^{1/2}\phi_\lambda\rangle_2
+\Gamma_{\vs,\alpha}(\lambda)\,\bar q_\psi q_\phi\,.
\end{align*}
It is then straightforward to check that $F_{\vs,\alpha}$ is
closed and bounded from below. Thus $F_{\vs,\alpha}$ is the quadratic
form associated with $H_{\vs,\alpha}$. Since 
$$\G_\lambda^\vs-\G_\lambda\in H^1(\RE^3)$$
one obtains $D(F_{\vs})\equiv D(F_0)$. Re-writing 
the quadratic form above by using the decomposition
entering in the definition of $D(F_0)$ and noticing that 
$$
\forall \psi\in H^1(\RE^3)\qquad (F_\vs+\lambda)
(\G_\lambda^\vs-\G_\lambda,\psi)=\langle\G_\lambda,iL_\vs\psi\rangle_2
$$
one obtains
\begin{align*}
&(F_{\vs,\alpha}+\lambda)(\psi,\phi)=\langle\nabla\psi_\lambda,\nabla\phi_\lambda\rangle_2+\lambda\langle\psi_\lambda,\phi_\lambda\rangle_2
+Q_\vs(\psi,\phi)\\&+
\left(\Gamma_{\vs,\alpha}(\lambda)+
(F_{\vs,\alpha}+\lambda)(\G^\vs_\lambda-\G_\lambda,\G^\vs_\lambda-\G_\lambda)
\right)
\bar q_\psi q_\phi\,,
\end{align*}
Since $L_\vs$ is skew-adjoint one has $\langle L_\vs\psi,\psi\rangle_2=0$ for
any real-valued $\psi\in H^1(\RE^3)$. Thus, by taking a real-valued $J_\epsilon\in C_0^\infty(\RE^3)$ such that 
$J_\epsilon$ weakly converges to the Dirac mass at the origin as
$\epsilon\downarrow 0$, one obtains (here $*$ denotes convolution)
\begin{align*}
&(F_{\vs,\alpha}+\lambda)(\G^\vs_\lambda-\G_\lambda,\G^\vs_\lambda-\G_\lambda)=
\langle\G_\lambda,iL_\vs(\G^\vs_\lambda-\G_\lambda)\rangle_2\\
=&\lim_{\epsilon\downarrow 0}\,\langle iL_\vs\G_\lambda*J_\epsilon,
\G^\vs_\lambda-\G_\lambda*J_\epsilon\rangle_2=
\lim_{\epsilon\downarrow 0}\,\langle iL_\vs\G_\lambda*J_\epsilon,
\G^\vs_\lambda\rangle_2=\\
&=\lim_{\epsilon\downarrow 0}\,\langle
(H_\vs+\lambda)(\G^\vs_\lambda-\G_\lambda)
*J_\epsilon,\G^\vs_\lambda\rangle_2\\
=&\lim_{\epsilon\downarrow 0}\,\langle
\G^\vs_\lambda-\G_\lambda,(H_\vs+\lambda)
G_\lambda^\vs*J_\epsilon,\rangle_2=
\lim_{\epsilon\downarrow 0}\,\langle
\G^\vs_\lambda-\G_\lambda,J_\epsilon,\rangle_2=\\
=&(\G^\vs_\lambda-\G_\lambda)(0)=\frac{\sqrt\lambda-\sqrt{\lambda-|\vs|^2/4}}{4\pi}\equiv\Gamma_\alpha(\lambda)-\Gamma_{\vs,\alpha}(\lambda)
\end{align*}
and the proof is done.
\end{proof}
\begin{remark} \label{remark} Let $J_\epsilon$ be a real-valued, compactly supported smooth
function weaky converging to the Dirac mass a zero  as $\epsilon\downarrow 0$.
For any $\psi=\psi_\lambda+q_\psi\G_\lambda$ and 
$\phi=\phi_\lambda+q_\phi\G_\lambda$, let us define 
$\psi_\epsilon:=\psi_\lambda+q_\psi\G_\lambda*J_\epsilon$ and 
$\phi_\epsilon:=\phi_\lambda+q_\phi\G_\lambda*J_\epsilon$. Then, since
$L_\vs$ is
skew-adjoint, one has
\begin{align*}
&\lim_{\epsilon\downarrow 0}\,\langle iL_\vs
\psi_\epsilon,\phi_\epsilon
\rangle_2=\lim_{\epsilon\downarrow 0}\,\left(
\langle iL_\vs\psi_\lambda,\phi_\lambda\rangle_2+
\bar q_\psi\langle \G_\lambda*J_\epsilon,iL_\vs\phi_\lambda\rangle_2\right.
\\&\left.+q_\phi\langle iL_\vs\psi_\lambda,\G_\lambda*J_\epsilon\rangle_2
-i\bar q_\psi q_\phi
\langle L_\vs\G_\lambda*J_\epsilon,\G_\lambda*J_\epsilon\rangle_2\right)\\
=&\lim_{\epsilon\downarrow 0}\,
Q_\vs(\psi_\epsilon,\phi_\epsilon)=Q_\vs(\psi,\phi)\,.
\end{align*}
Thus $Q_\vs$ is the natural extension 
to $D(F_0)$ of the quadratic form associated with $iL_\vs$.
\end{remark}

\end{section}
\begin{section}{The Schr\"odinger equation with a moving point interaction}
Let us now consider a differentiable curve $y:\RE\to\RE^3$ and put
$\vs(t)\equiv \frac{dy}{dt}(t)$. Thus one has the families of
self-adjoint operators and associated quadratic forms
$$
H_{\alpha,y}(t):D(H_{\alpha,y}(t))\to L^2(\RE^3)\,,
$$
$$
F_{\alpha,y}(t):D(F_y(t))\times D(F_y(t))\to\RE\,,
$$
$$
H_{\vs,\alpha}(t):D(H_{\alpha,y}(t))\to L^2(\RE^3)\,,
$$
$$
F_{\vs,\alpha}(t):D(F_0)\times D(F_0)\to\RE\,.
$$ 
Now we can state our main result:
\begin{theorem}\label{theo} Let $y\in C^2(\RE;\RE^3)$. 
Then there is a unique strongly continuous unitary propagator 
$$U_{t,s}:L^2(\RE^3)\to L^2(\RE^3)\,,\qquad (t,s)\in\RE^2\,,$$
such that \p
1) $$U_{t,s}D(F_{\alpha,y}(s))=D(F_{\alpha,y}(t))\,;
$$
2) each $U_{t,s}$ is strongly continuous as a map from $D(F_{y}(s))$ 
onto $D(F_{y}(t))$ with respect to the Banach topologies
induced by the bounded from below closed quadratic forms $F_{\alpha,y}(s)$ and
$F_{\alpha,y}(t)$ respectively; \p
3) $$\forall\psi\in D(F_y(s))\,,\qquad 
t\mapsto F_{\alpha,y}(t)(U_{t,s}\psi,U_{t,s}\psi)\qquad \text{\it is 
in}\quad C(\RE;\RE)\,;
$$
4)$$
\forall\psi\in D(F_y(s))\,,\qquad 
t\mapsto U_{t,s}\psi\qquad \text{\it is 
in}\quad C^{1}\left(\RE;D(F_y(\cdot))^*\right)\,,
$$
where $D(F_y(t))^*$ denotes the dual of $D(F_y(t))$ with respect to
the $L^2(\RE^3)$ scalar product;\p
5)$$
\forall\psi\in D(F_y(s))\,,\ \forall\phi\in D(F_y(t))\,,
\quad \left(i\,\frac{d}{dt}\,U_{t,s}\psi,\phi\right)_t
=F_{\alpha,y}(t)(U_{t,s}\psi,\phi)\,,
$$
where $(\cdot,\cdot)_t$ denotes the duality between $D(F_y(t))$ and $D(F_y(t))^*$.\p
If $y\in C^3(\RE;\RE^3)$ then \p
6)
$$U_{t,s}\tilde D(H_{\alpha,y}(s))= \tilde D(H_{\alpha,y}(t))\,,$$ 
where $$\tilde D(H_{\alpha,y}(t)):=V_t D(H_{\alpha,y}(t))\,,\quad
V_t\psi(x):=e^{i\vs(t)\cdot x/2}\psi(x)\,.$$ 
\end{theorem}
\begin{proof} By Theorem \ref{form} we have that $y\in C^2(\RE;\RE^3)$
implies that
$$
\forall\,\psi,\phi\in D(F_0)\,,\qquad  
t\mapsto F_{\vs,\alpha}(t)(\psi,\phi)\qquad \text{\rm is in}\quad 
C^{1}(\RE)\,.
$$
Let $T>0$. By Kisy\'nski's theorem (see the Appendix) applied to the family of strictly positive self-adjoint operators 
$$H_{\vs,\alpha}(t)+\lambda\,, \quad t\in [-T,T]\,,\quad\lambda> 
(4\pi\min(0,\alpha))^2+\frac{1}{4}\,\sup_{t\in
[-T,T]}|\vs(t)|\,,$$ one knows that $H_{\vs,\alpha}(t)$, $t\in[-T,T]$, 
generates a strongly
continuous unitary propagator $\tilde U^T_{t,s}$, $(s,t)\in
[-T,T]^2$. By unicity if $T'>T$ then $\tilde U^T_{s,t}=U^{T'}_{s,t}$ for any
$(s,t)\in [-T,T]^2\subset [-T',T']^2$. Thus we obtain an unique 
strongly continuous unitary propagator 
$\tilde U_{t,s}$, $(s,t)\in\RE^2$, generated by the
family $H_{\vs,\alpha}(t)$, $t\in\RE$. Such a propagator is also a strongly
continuous propagator on $D(F_0)$ with respect to the Banach topology induced 
by the bounded from below closed quadratic form $F_{\alpha,0}$.\par
Considering the
unitary map
$$
T_{t}:L^2(\RE^3)\to L^2(\RE^3)\,,\qquad T_{t}\psi(x)
:=\psi(x+y(t))\,,
$$  
we define then the strongly continuous unitary propagator
$$
U_{t,s}:=T_t^{-1}\tilde U_{t,s}T_s\,.
$$
Since $T_t$ is a bounded operator from $D(F_y(t))$ onto $D(F_0)$ one has that 
$U_{t,s}$ is a bounded operator from $D(F_{y}(s))$ 
onto $D(F_{y}(t))$ with respect to the Banach topologies
induced by the bounded from below closed quadratic forms $F_{\alpha,y}(s)$ and
$F_{\alpha,y}(t)$ respectively. Moreover,
for all $\psi\in D(F_y(s))$, the map 
$$
t\mapsto F_{\alpha,y}(t)(U_{t,s}\psi,U_{t,s}\psi)\equiv 
F_{\alpha,0}(\tilde U_{t,s}T_s\psi,\tilde U_{t,s}T_s\psi)
$$
is continuous. Let us now show that, 
for all $\psi\in D(F_y(s))$ and for all $\phi\in D(F_y(t))$
one has
$$
\left(i\,\frac{d}{dt}\,U_{t,s}\psi,\phi\right)_t=F_{\alpha,y}(t)
(U_{t,s}\psi,\phi)\,.
$$
For any $\psi\in D(F_y(s))$ and $\phi\in D(F_y(t))$ there exist
$\tilde\psi$ and $\tilde\phi$ in $D(F_0)$ such that $T^{-1}_s\tilde\psi=\psi$ and 
$T^{-1}_t\tilde\phi=\phi$. Thus equivalently we need to show that
$$
\left(i\,\frac{d}{dt}\,T_t^{-1}\tilde U_{t,s}\tilde\psi,T^{-1}_t\tilde\phi\right)_t=
F_{\alpha,y}(t)(T^{-1}_t\tilde U_{t,s}\tilde\psi,T^{-1}_t\tilde\phi)
\equiv F_{\alpha,0}(\tilde U_{t,s}\tilde\psi,\tilde\phi)\,.
$$
Since
\begin{align*}
&\left(i\,\frac{d}{dt}\,T_t^{-1}\tilde
U_{t,s}\tilde\psi,T^{-1}_t\tilde\phi\right)_t
=\left(i\,T_t\,\frac{d}{dt}\,T_t^{-1}\tilde
U_{t,s}\tilde\psi,\tilde\phi\right)\\=&
\left(i\,T_t\left(\,\frac{d}{dt}\,T_t^{-1}\right)
\tilde U_{t,s}\tilde\psi,\tilde\phi\right)+
\left(\,i\,\frac{d}{dt}\,\tilde U_{t,s}\tilde\psi,\tilde\phi\right)\\=&
\left(i\,T_t\left(\,\frac{d}{dt}\,T_t^{-1}\right)
\tilde U_{t,s}\tilde\psi,\tilde\phi\right)
+F_{\vs,\alpha}(t)(\tilde U_{t,s}\tilde\psi,\tilde\phi)\,,
\end{align*}
by Theorem \ref{form} we need to show that 
$$
\left(i\,T_t\,\frac{d}{dt}\,T_t^{-1}
\tilde\psi,\tilde\phi\right)=-Q_\vs(\tilde\psi,\tilde\phi)\,.
$$
This is obviously true in the case either $\tilde\psi$ or $\tilde\phi$
is in $H^1(\RE^3)$ and, by taking $J_\epsilon$ as in Remark
\ref{remark}, 
\begin{align*}
&\left(T_t \,\frac{d}{dt}\,T_t^{-1}
\G_\lambda,\G_\lambda\right)
=\lim_{\epsilon\downarrow 0}\,
\left\langle T_t \,\frac{d}{dt}\,T_t^{-1}
\G_\lambda*J_\epsilon,\G_\lambda*J_\epsilon\right\rangle_2\\
=&-\lim_{\epsilon\downarrow 0}\,\langle L_\vs \G_\lambda*J_\epsilon,\G_\lambda*J_\epsilon\rangle_2=0\,.
\end{align*}
Thus point 5) is proven. Point 6) follows from Kisy\'nski's theorem 
again by noticing that if $y\in C^3(\RE;\RE^3)$ then $\tilde U_{t,s}$
maps $D(H_{\vs,\alpha}(s))$ onto $D(H_{\vs,\alpha}(t))$ and that 
$$
D(H_{\vs,\alpha}(t))\equiv T_tV_t D(H_{\alpha,y}(t))\,.
$$
\end{proof}

\end{section}
\begin{section}{Appendix: the Kisy\'nski's theorem}
For the reader's convenience in this appendix we recall 
Kisy\'nski's theorem. For the proof we refer 
to Kisy\'nski's original paper \cite{[K]} (see in particular
\cite{[K]}, section 8. Also see \cite{[S]}, section II.7). \par
Let us remind that the double family $U_{t,s}$, $(t,s)\in
[T_1,T_2]^2$, 
is said to be a strongly continuous 
propagator on the Hilbert space $\H$ if each $U_{t,s}$ is a bounded
operator on $\H$, the map $(t,s)\mapsto U_{t,s}$ 
is strongly continuous, $U_{s,s}={\mathsf 1}$ and the
Chapman-Kolmogorov equation
$$
U_{t,r}U_{r,s}=U_{t,s}
$$
holds. Such a propagator is then said to be unitary if each 
$U_{t,s}$ is unitary.
\begin{theorem} Let $A(t)$, $t\in [T_1,T_2]$, be a family of 
strictly positive self-adjoint operators
on the Hilbert space $\left(\H,\langle\cdot ,\cdot \rangle\right)$ 
with time-independent
form domain $\H_+$. Suppose that 
$$\forall\psi,\phi\in\H_+\qquad  
t\mapsto F(t)(\psi,\phi)\qquad \text{\it is in}\quad 
C^k\left([T_1,T_2];\RE\right)\,,
$$
where $F(t)$ denotes the quadratic form associated with $A(t)$. 
\p If $k=1$ then there is a unique strongly continuous unitary propagator 
$$U_{t,s}:\H\to\H\,,\qquad (s,t)\in [T_1,T_2]^2\,,$$
such that \p
1) $$U_{t,s}\H_+=\H_+\,;
$$
2) $U_{t,s}$ is a strongly continuous propagator on 
$\left(\H_+,\langle\cdot ,\cdot \rangle_+\right)$, where
$\langle\cdot ,\cdot \rangle_+$ denotes any of the equivalent 
scalar products  
$$\langle\psi,\phi\rangle_{t,+}:=F(t)(\psi ,\phi)\,;$$
3)$$
\forall\psi\in\H_+\qquad 
t\mapsto U_{t,s}\psi\qquad \text{\it is 
in}\quad C^{1}\left([T_1,T_2];\H_-\right)\,,
$$
where $\left(\H_-,\langle\cdot ,\cdot \rangle_-\right)$ is
the completion of $\H$ endowed with any of 
the equivalent scalar products 
$$\langle\psi,\phi\rangle_{t,-}:=\langle A(t)^{-1/2}\psi ,A(t)^{-1/2}\phi\rangle\,;$$
4)$$
\forall\psi,\phi\in\H_+\qquad \left(i\,\frac{d}{dt}\,U_{t,s}\psi,\phi\right)
=F(t)(U_{t,s}\psi,\phi)\,,
$$
where $(\cdot,\cdot)$ denotes the duality between $\H_+$ and $\H_-$. 
If $k=2$ then \p
5)$$
U_{t,s}\,D(A(s))=D(A(t))\,,
$$ 
6)$$
\forall\psi\in D(A(s))\quad 
t\mapsto U_{t,s}\psi\quad \text{\it is 
in}\quad C^1([T_1,T_2];\H)\cap C([T_1,T_2];D(A(\cdot)))\,,
$$
7) 
$$
i\,\frac{d}{dt}\,U_{t,s}\psi=A(t)U_{t,s}\psi\,.
$$
\end{theorem}
\end{section}

\end{document}